\documentclass[achesmo,epsfig,floats,twocolumn,superscriptaddress,amssymb,amsmath,floatfix]{article}
\usepackage{graphicx}
\usepackage{bm}  
\usepackage{amssymb}
\usepackage{color}
\usepackage{amscd}
\usepackage{epsfig}

\title{Designing phoretic micro- and nano-swimmers}

\author{R. Golestanian$^1\footnote{r.golestanian@sheffield.ac.uk}$,
T.B. Liverpool$^2$, A. Ajdari$^{3,4}\footnote{armand.ajdari@espci.fr}$
\\
\normalsize{\em $^1$ Department of Physics and Astronomy, University
of
Sheffield, Sheffield S3 7RH, UK}\\
\normalsize{\em $^2$ Department of Applied Mathematics, University
of Leeds, Leeds
LS2 9JT, UK}\\
\normalsize{\em $^3$ Physico-Chimie Th\'eorique, UMR CNRS 7083,
ESPCI, 10 rue
Vauquelin, 75005 Paris, France}\\
\normalsize{\em $^4$ DEAS, Harvard University, 29 Oxford Street,
Cambridge MA 02138, U.S.A}}

\date{\today}

\begin{document}
\maketitle
\begin{abstract}
Small objects can swim by generating around them fields or gradients
which in turn induce fluid motion past their surface by phoretic
surface effects. We quantify for arbitrary swimmer shapes and
surface patterns, how efficient swimming requires both surface
``activity'' to generate the fields, and surface ``phoretic
mobility.'' We show in particular that (i) swimming requires
symmetry breaking in either or both of the patterns of "activity"
and ``mobility,'' and (ii) for a given geometrical shape and surface
pattern, the swimming velocity is size-independent. In addition, for
given available surface properties, our calculation framework
provides a guide for optimizing the design of swimmers.
\end{abstract}

\paragraph{Introduction.} In the current miniaturization race towards small
motors and engines, a rapidly expanding subdomain is the quest for
autonomous swimmers, able to move in fluids which appear very
viscous given the small length scales (low Reynolds number). Robotic
microswimmers that generate surface distortions is an avenue (e.g.
by mimicking sperms \cite{Dreyfus}), but it seems equally
interesting to try to take advantage of physical phenomena that
become predominant at small scales. Interfacial "phoretic" effects
(electrophoresis, thermophoresis, diffusiophoresis, \cite{Anderson})
are from this standpoint a natural avenue given the increased
surface to volume ratio.

Very recently, in line with earlier suggestions of
self-electrophoresis of objects that would be able to generate an
electric field around them \cite{Mitchell,Anderson,Lammert} and our
recent proposal of self-diffusiophoresis \cite{gla-lett}, many
experimental reports have appeared of heterogeneous objects (e.g.
rods of hundreds of nanometers) swimming using different catalytic
or reactive chemistry at their different ends
\cite{Paxton,Mano,Ozin,Vicario}. Although many mechanisms may
contribute to the observed motion \cite{Paxton}, phoretic mechanisms
have been shown to be essential for some systems \cite{Paxtonpriv}.

In this Letter we consider generally self-phoretic motion, where
the swimmer, generates through its surface activity
gradients of (at least) a quantity (concentration of a
dissolved species, electric potential, temperature), which in turn
induce motion through classical interfacial phoretic processes.
Because it is now possible to fabricate such objects with controlled
shape and patterns of surface properties  \cite{Paxton,Ozin,Perro},
we discuss how these design parameters affect the swimmer's
velocity. A general framework is proposed to compute this velocity
for arbitrary swimmer size, shape and surface pattern. We then work
out explicitly a few examples for spheres and rods, geometries
amenable to analytical calculations, and corresponding to existing
means of fabrication \cite{Paxton,Ozin}. From these examples we
extract generic rules as to the role of the pattern symmetry, of the
swimmer size and shape, before concluding with a few remarks.

\paragraph{General formalism.}
We start with a reminder of what phoretic mechanisms are
\cite{Anderson}. Consider e.g. a colloid in a solution where the
concentration of a solute  $c$ exhibits a gradient. The colloid is
then set into motion due to the induced imbalance of osmotic effects
within the solid/liquid interfacial structure at the surface of the
object.
When the thickness of the interfacial layer is thin compared to the object,
the resulting flow is most
conveniently described by an effective slip velocity of the
liquid past the solid, proportional to the local gradient of $c$ \cite{Anderson}:
\begin{equation}
{\mathbf v}_s({\bf r}_s)= \mu({\bf r}_s) ({\mathbf {I - nn}}) \cdot \boldsymbol{\nabla}
c({\bf r}_s)
\end{equation}
where ${\mathbf n}$ is the local normal to the surface,
$\mathbf {I}$ is the identity tensor. $\mu({\bf r}_s)$
is the local "surface phoretic mobility", positive or negative
depending on specifics of the solute/surface interactions \cite{Anderson}.

This distribution of slip on the object's surface induces a net drift velocity ${\mathbf V}$,
which can be computed using the reciprocal
theorem for low Reynolds-number hydrodynamics \cite{Stone}.
Its components in a referential (${{\hat{\mathbf e}}_i}$) are:
\begin{equation}
{\mathbf V} \cdot {\hat{\mathbf e}}_i= - \int \!\!\!\!\! \int_{\rm S} d {\bf r}_s \,\,
{\mathbf{n} \cdot  \boldsymbol{\sigma}}_i \cdot {\bf v}_s
\label{gen_swim_vel}
\end{equation}
where $ \boldsymbol{\sigma}_i$ is the hydrodynamic stress tensor at the surface of
an object of similar shape
dragged by an applied unit force exactly equal to
${\hat{\mathbf e}}_i$, in the absence of slip.

Similar principles describe the motion of the object in a gradient
of electric potential (electrophoresis), or in a temperature
gradient (thermophoresis). The above formalism can be applied upon
simple replacement  of $c$ by the potential $\phi_{el}$ or by the
temperature $T$. For electrophoresis $\mu$ is up to a constant the
so-called zeta-potential \cite{Anderson}.
For all these situations, achieving large values of $\mu$ requires
engineering of the surface in relation to the solvent (e.g.
nanometer scale roughness and solvophibicity do matter
\cite{Netz,Ajdari}).

We now focus on objects that generate the gradients {\em
themselves}, e.g. using the chemical free-energy available in the
solvent \cite{Paxton,Ozin}. Returning to self-diffusiophoresis, the
simplest situation corresponds to a solute generated by active
sources and sinks on the surface of the object, and diffusing in the
bulk. At steady-state, in the reference frame of the object, and
neglecting distortions induced by the flow (small Peclet number),
the solute concentration in the liquid is given by:
\begin{equation}
D \nabla^2 c =0,\label{c-eq-1}
\end{equation}
\begin{equation}
-D \, \mathbf{n} \cdot \boldsymbol{\nabla} c({\bf r}_s) =\alpha({\bf
r}_s),\label{c-eq-BC}
\end{equation}
where $D$ is the diffusion coefficient, and $\alpha({\bf r}_s)$
measures the ``surface activity'' at position ${\bf r}_s$ on the
surface, i.e. the generation or consumption of solute by a chemical
reaction. In general, describing this process involves additional
coupled transport problems for other species involved in the surface
reactions. For simplicity, we nevertheless proceed with a fixed
local value of $\alpha$. This should well describe situations where
the solvent contains an excess of the reactants necessary to produce
and destroy the solute of interest, so that the density of active
sites on the surface is what limits the fluxes.  $\alpha$ is then
this density times the rate of the chemical reaction per site
\cite{gla-lett}.

For self-electrophoresis, similar formulae hold upon replacing $c$
by $\phi_{el}$ and $D$ by the solution electrical conductivity.
$\alpha$ is then the electrical current injected by the surface.
Existing realizations \cite{Mano,Paxtonpriv}, correspond to no net
current generation $\int\!\!\! \int_{\rm S} d{\bf r}_s \;\alpha({\bf
r}_s)=0$, with the ions produced on one side of the swimmer and
consumed on the other one, while electrons are transported through
the body of the swimmer. Similarly, for thermophoresis, beyond
replacing $c$ by $T$ and $D$ by the thermal conductivity, the
boundary condition (4) must be adjusted to account for heat
transport through the object.

The general formal description (1-4) predicts a swimming velocity
$V\sim \alpha \mu /D$, linear in the surface properties $\alpha$ and
$\mu$, and inversely proportional to the medium conductivity $D$.
More interestingly, once the pattern of surface properties and the
object shape are given, $V$ is independent of the size $R$ of the
swimmer.
 This central point emphasizes the robustness
of this swimming strategy against downsizing.
In contrast, an external body force on the object
scaling as $F\sim R^3$ (dielectrophoresis, magnetophoresis)
leads in this viscous realm
to velocities $\sim R^2$, which decrease rapidly with length scale.
Further, our approach (1-4) allows us to relate quantitatively the velocity $V$
to the design of the swimmer
(shape, surface patterns of $\alpha$ and $\mu$),
as we show analytically for spheres and rods.

\paragraph{Spheres.}

For a sphere of radius $R$, as ${\mathbf n \cdot}  \boldsymbol{\sigma}_i=(1/4\pi
R^2){\hat{\mathbf e}}_i$, equation (2) becomes
\begin{equation}
{\mathbf V}= - \frac{1}{4\pi R^2} \int \!\!\! \!\! \int_{\rm S} d{\bf r}_s \,\,
\mu({\bf r}_s) ({\mathbf {I - nn}}) \cdot \boldsymbol{\nabla} c ({\bf r}_s),\label{V-main}
\end{equation}
We focus further on azimuthally symmetric patterns, with surface
quantities depending only on the ``latitude'' angle $\theta$ with
the polar axis ${{\hat{\mathbf e}}_z}$. We expand in Legendre
polynomials. The surface activity
$\alpha(\theta)=\sum_{\ell=0}^{\infty} \alpha_{\ell} P_{\ell}(\cos
\theta)$, generates a concentration variation given by Eqs.
(\ref{c-eq-1}) and (\ref{c-eq-BC}):
\begin{equation}
c(r,\theta)=c_{\infty}+\frac{R}{D} \sum_{\ell=0}^{\infty}
\frac{\alpha_{\ell}}{\ell+1} \left(\frac{R}{r}\right)^{\ell+1}
P_{\ell}(\cos \theta),\label{c-sol-1}
\end{equation}
where $c_{\infty}$ is the concentration at infinity. For a surface mobility
$\mu(\theta)=\mu_{\ell} P_{\ell}(\cos \theta) \label{mu-ell} $,
the corresponding gradient yields a swimming velocity (5):
\begin{equation}
{\bf V}=-\frac{{{\hat{\mathbf e}}_z}}{D} \sum_{\ell=0}^{\infty}
\left(\frac{\ell+1}{2 \ell+3}\right) \; \alpha_{\ell+1}
\left[\frac{\mu_{\ell}}{2 \ell+1}-\frac{\mu_{\ell+2}}{2
\ell+5}\right]. \label{v-formula-1}
\end{equation}

As a first check, we recover
\cite{gla-lett} that a single active site at the pole
$\alpha= \tau^{-1} \delta_{2D}({\bf r}_s)$ of a sphere of uniform mobility $\mu$
yields $V=\frac{\mu}{4 \pi R^2 D \tau} $. By linearity,
an emitter at the back pole with rate $1/\tau_e$ and a consumer at
the front pole with rate $1/\tau_c$, results in
$ V=\frac{\mu}{4 \pi R^2 D}
\left(\frac{1}{\tau_e}+\frac{1}{\tau_c}\right)$.

\begin{figure}[t]
\includegraphics[width=1.00\columnwidth]{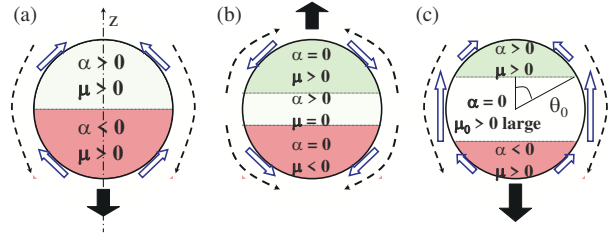}
\caption{Three spherical  swimmers described in the text. The flux
of $c$ (opposite to the gradient) is depicted with the dashed
arrows, while the resulting slip velocity of the liquid along the
sphere is described by open arrows. The net swimming direction is
given by the thick dark arrow. (a) a Janus swimmer, with a
homogeneous $\mu$ and a broken top-down symmetry in $\alpha$: one
hemisphere produces $c$ while the other consumes it.  (b) a Saturn
swimmer with an equatorial belt producing $c$, and a broken symmetry
in the phoretic mobility between the two hemispheres (here even
opposite signs), which induce swimming in the same direction from
the (opposite) equator to poles gradients. (c) a three-slice design
in with polar slices efficient in producing and consuming $c$, while
the equatorial slice is chosen for its large mobility. Properly
choosing $\theta_0$ permits maximization of the swimming velocity
(Fig. 2)} \label{fig:three-spheres}
\end{figure}

To access useful general principles we now focus on three simple
designs (Figure 1), within reach of current fabrication techniques
\cite{Perro}. The first is the Janus sphere, with two hemispheres
covered with a different enzymatic or catalytic material.
Generically, their mobilities $\mu$ also differ, so we take:
\begin{equation}
\left(\alpha(\theta),\mu(\theta)\right)=\left\{\begin{array}{ll}
(\alpha_{+},\mu_{+}), & \; 0< \theta <\frac{\pi}{2},  \\ \\
(\alpha_{-},\mu_{-}), & \; \frac{\pi}{2}<\theta<\pi,
\end{array} \right. \label{alpha-janus}
\end{equation}
and find analytically
\begin{equation}
{\bf V}=\frac{1}{8} \;\frac{1}{D} (\alpha_{-}-\alpha_{+})
(\mu_{+}+\mu_{-})\, {{\hat{\mathbf e}}_z}.
\end{equation}
This shows that symmetry breaking in the chemical activity ($
\alpha_{-} \neq \alpha_{+} $) is essential to propulsion, and that
the swimming velocity is larger if the two mobilities are of the
same sign (which dictates the swimming direction).

The second design is a ``Saturn'' particle, where the surface
activity is concentrated around the equator, and where symmetry is
now broken by choosing hemispheres of different mobilities. For
computational convenience we solve the very similar problem defined
by:
\begin{equation}
\alpha(\theta)=\alpha_0 (1-\cos ^2 \theta),\,\,\,
\mu(\theta)=\mu_0 \cos \theta, \label{saturn}
\end{equation}
and find
\begin{equation}
{\bf V}=\frac{4}{45} \;\frac{\mu_0 \alpha_0}{D} \, {{\hat{\mathbf e}}_z}.
\end{equation}
So symmetry breaking in mobility alone also leads to swimming:
the activity at the equator generates gradients of $c$ towards
the poles that generate slip velocities in the same direction
along $z$ as the sign of $\mu$ is opposite on the two hemispheres (Figure 1b).

The third design is that of a three-slice sphere,
that illustrates typical considerations that arise
when trying to increase by design the swimming velocity.
From the scaling analysis above one should obviously aim for
surfaces with both large values of $\alpha$ and $\mu$.
However, in many cases these goals will not be met by a single surface,
so one may instead look for ways to combine in an efficient design surfaces with
large $\alpha$ and surfaces with large $\mu$.
As an example, let us try to upgrade the Janus design of Figure 1a
by adding a passive belt of large phoretic mobility $\mu_0$, which
leads to the swimmer of Figure 1c:
\begin{equation}
\left(\alpha(\theta),\mu(\theta)\right)=\left\{\begin{array}{ll}
(\alpha_{+},\mu_{+}), & \; 0< \theta <\theta_+,  \\ \\
(0,\mu_{0}), & \; \theta_+< \theta <\pi-\theta_-,  \\ \\
(\alpha_{-},\mu_{-}), & \; \pi-\theta_-<\theta<\pi,
\end{array} \right. \label{alpha-three-slice}
\end{equation}
To reach the largest swimming velocity for given surface properties,
one can adjust the values of $\theta_+$ and $\theta_-$ in a
compromise between increasing the mobility by enlarging the belt
without shrinking too much the poles that generate the gradient.
Figure 2 shows results for the model case
$\alpha_+=-\alpha_-=\alpha$, $\mu_+=\mu_-=\mu$ of same sign than
$\mu_0$, $\theta_+=\theta_-=\theta_0$: for $\mu_0>\mu$ there is a
value of $\theta_0$ that maximizes the speed, which is always larger
than $\pi/3$, i.e. no matter how large $\mu_0$, the "optimal" belt
should not be made larger than a fixed value. The relative flatness
of the bottom curve in Figure 2 shows a moderate dependence on
$\theta_0$ around this value, so that the swimming velocity should
be rather robust to small imprecisions in the fabrication.

\begin{figure}[t]
\includegraphics[width=1.00\columnwidth]{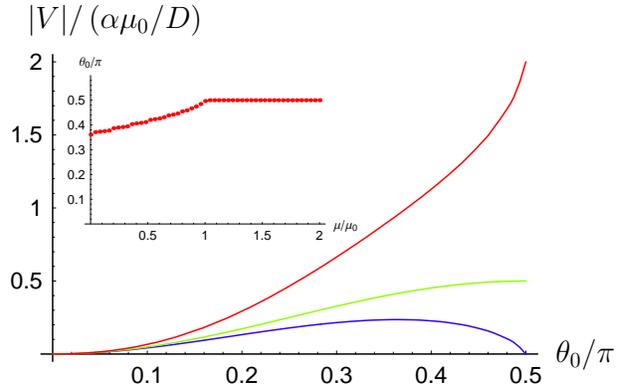}
\caption{Reduced swimming velocity $V/(\alpha \mu_0/D)$ of the
three-slice swimmer of Figure 1c, as a function of "polar" angle
$\theta_0$ for various values of $\mu/\mu_0$: $0$, $1$, $4$ (from
bottom to top). The optimal value at low $\mu/\mu_0$ reflects the
compromise between large poles to have strong chemical gradients,
and a large equatorial belt for these gradients to translate into
large flow motion. Inset: optimal angle $\theta_0$ against
$\mu/\mu_0$,
. }
\label{fig:optimal_angle}
\end{figure}

\paragraph{Thin rods.} Because of the many experimental realizations of such
systems \cite{Paxton, Ozin}, we discuss the case of long cylindrical
rods ( length $2L$ and radius $b\ll L$) with surface properties
varying only along their axis $-L \le z \le L$, i.e. $\mu = \mu (z),
\alpha = \alpha(z)$.
In the limit $L \gg b$ slender body theory can be used, and
remarkably equation (\ref{gen_swim_vel}) yields a simple form for
the swimming velocity along $z$:
\begin{equation}
V \simeq -\frac{1}{2L} \int_{-L}^{L} d
z \; \mu(z) \partial_{z} c_s,\label{SS-2}
\end{equation}
In the same limit,
the surface concentration $c_s$
is related to the surface activity $\alpha(z)$
by \cite{Paxton}
\begin{equation}
c_s(z) \simeq \frac{b}{2 D} \int_{-L}^{L} d z' \;
\frac{\alpha(z')}{|z-z'|}.
\label{c-rod-sol-1}
\end{equation}
These two equations give the velocity of a rod for any "bar-coded"
surface properties $\alpha$ and $\mu$. For a Janus rod
\begin{eqnarray}
\left(\alpha(z),\mu(z)\right)&=&\left\{\begin{array}{ll}
(\alpha_{-},\mu_{-}), & \; -L< z < 0,  \\ \\
(\alpha_{+},\mu_{+}), & \; 0<z<L,
\end{array} \right. \label{alpha-rod-janus}
\end{eqnarray}
we find
\begin{eqnarray}
V = {1 \over 4 D} \left({b \over L}\right) \ln \left({L \over
4b}\right)  \left( \mu_{+}\alpha_{-} -  \mu_{-}\alpha_{+}\right)
\end{eqnarray}
which illustrates a few important points:
(i) again swimming is obtained by symmetry breaking
in the chemical activity or/and in the phoretic mobility,
(ii) swimming with uniform activity ($\alpha=$constant) is possible
- in contrast to the special case of spheres for which this leads to
no tangential gradients of $c$ -, (iii) the velocity scale is
decreased by a factor of $\sim b/L$ when compared to a sphere of
radius $\sim L$, because the surface generated flux of $c$ is
reduced by a similar factor, and (iv) $V$ is up to logarithms
proportional to $1/L$ at fixed rod diameter, but scale independent
if length and diameter are varied in the same proportion.

\paragraph{Discussion.}

The swimming velocity $V$
of a phoretic swimmer, is independent of its size and
scales as $V \sim  \alpha \mu/D$, where $\alpha$ and $\mu$ are some averages
quantifying the surface
activity generating the gradient and the surface phoretic mobility.
The exact value of $V$ depends on the shape and surface pattern
in a computable way. For example, spheres swim faster
than rods of same longest dimension, and improvement
can be obtained by properly combining various types of surfaces.

- This type of formalism holds for self-diffusiophoresis (with
$\alpha$ the surface reaction density, $\mu$ the diffusiophoretic
mobility, and $D$ the solute diffusion coefficient), for
self-electrophoresis (with $\alpha$ the electrical current density
injected in the solution, $\mu$ the electrophoretic mobility, and
$D$ the solution electrical conductivity), and for
self-thermophoresis (with $\alpha$ the surface heat flux injected in
the solution, $\mu$ the thermophoretic mobility, and $D$ the
solution thermal conductivity). As for motion driven by external
gradients, the largest velocities are expected for highly charged
surfaces at low ionic strengths, which yield large mobilities $\mu$
because of the thick interfacial electrical double-layer.


- The present line of reasoning can be adapted to
more accurate descriptions of surface chemistry
for specific experimental conditions.
What is required in a given situation is an appropriate scheme for computing
$\alpha({\bf r}_s)$. For example if $c$ is produced at the surface from some
chemical at concentration
$c_{fuel}^{\infty}$ far away from the object, various regimes show up.
We have focused on kinetics limited by the number of catalytic sites on the surface,
but for low values of $c_{fuel}^{\infty}$ the reaction becomes diffusion limited
so that $\alpha$ depends on the global geometry.

- Due to thermal rotational diffusion,
if no external field orients the objects, the direction of motion persists
only for a time $\tau\simeq 1/D_{rot}$ , where $D_{rot}\simeq k_BT/\eta R^3$
and $R$ is the largest dimension of the object.
Note however that the field $c$ typically adjusts much faster to a change of orientation
($R^2/D \ll 1/D_{rot}$), so our steady-state estimates still hold for the
instantaneous velocity.
A simple Langevin dynamics with a diffusion coefficient $D_{trans}\sim k_BT/\eta R$
and drift at $V$ along a stochastically changing direction ($D_{rot}=1/\tau$) leads to
diffusive behaviour at long times ( $t \gg \tau$) with an activity-enhanced diffusion coefficient
$D_{eff}= D_{trans}+\frac{1}{6} V^2 \tau$.
This enhancement may be significant and measurable for not too small swimmers,
i.e. $R > (k_BT/\eta V)^{1/2}$, of order $\sim$ a few $100$~nm for $V\sim 10 ~\mu$m.s$^{-1}$ in water.

For nano-rods of radius $b$ fixed due to fabrication constraints
\cite{Paxton}, the scaling of the previous quantities become (using
the shorthand $V_0= \alpha / \mu D$ and forgetting logarithms)
$V\simeq V_0 (b/L)$, $\tau \sim \eta L^3 /k_BT$, and an
activity-induced enhancement of the bare translational coefficient
$D_{trans}\sim k_BT/\eta L$ by a quantity $\sim (\eta V_0^2/k_BT)
b^2 L$, significant if $L/b > (k_BT/\eta V_0b^2)$.

- Beyond the ``linear'' swimmers described here, our approach can
also describe active ``spinners,'' which rotate using the same
mechanisms \cite{Paxton,Ozin}. Their angular velocity can be
computed by replacing (4) by the formula for angular swimming speed
given in Ref. \cite{Stone}, and scales as $\Omega \sim \alpha \mu /D
R$, decreasing with object size for a given shape and geometrical
surface pattern.

- There is a growing interest in the collective behaviour of
swimming organisms (see e.g. \cite{Hernandez} and references
therein). Phoretic swimmers are special with regards to these
questions as they interact through both their hydrodynamic fields
(decaying as the inverse of the cube of their relative distance
$r^{-3}$) \cite{Anderson} and the $c$-field gradients they generate
(which decay as $r^{-2}$ for a net production/absorption of $c$ per
object, and as $r^{-3}$ otherwise).

In conclusion, we have provided generic considerations for the
design of small phoretic swimmers (size (in)dependence of the
swimming velocity, necessary symmetry breaking in the surface
pattern), as well as quantitative procedures to estimate the
influence of their velocity. Beyond these considerations, fast
motion of course relies on clever surface engineering to obtain
large surface ``activity'' and ``phoretic mobility.'' We hope that
our calculations will stimulate experimental studies of swimmers
with given surface chemistries but various surface patterns,
e.g. nanorods \cite{Paxton,Ozin} and spheres \cite{Perro,Howse},
keeping in mind that other effects (e.g. bubble generation
\cite{Ozin}) may compete with the ones described here.


\end{document}